\documentstyle[referee]{l-aa}
\input{epsf}
\input{rotate}
\begin{document}

\thesaurus{04(02.07.1; 11.11.1; 03.13.1; 12.04.1)}
\title{Rotation curves for spiral galaxies and  non-Newtonian
gravity:\hfill\break A phenomenological approach}

\author{C. Rodrigo-Blanco \and J.\ P\'erez-Mercader}

\institute{Laboratorio de Astrof\'{\i}sica Espacial y F\'{\i}sica
Fundamental (LAEFF),
P.O. Box 50727, E-28080 Madrid, Spain.}
\offprints{crb@laeff.esa.es}
\date{Received date, accepted date}

\maketitle

\begin{abstract}
Rotation curves of spiral galaxies are known with reasonable
precision for a large number of galaxies with similar morphologies.
The data implies that non-Keplerian fall--off  is seen. This
implies that (i) large amounts of dark matter must exist at
galactic scales or (ii) that Newtonian gravity must somehow be
corrected.
We present a method for inverting the integral relation between an
elemental law of gravity (such as Newton's) and the gravitational
field generated by a
thin disk distribution with exponential density.
This allows us to identify, directly from observations,
{\it extensions} of Newtonian gravity
 with the property of fitting a large class of
rotation curves. The modification is inferred from the observed
rotation curve and is finally written in terms of
Newton's constant or the effective potential of a test mass
moving in the field generated by a point-like particle.

\end{abstract}


When trying to understand the dynamics of large scale astrophysical
systems, for example galaxies, we assume that the dominant interaction
 at that scale is gravity. This implies the use of
Einstein's General theory of Relativity (GR), so that in the limit
when the speeds involved are much smaller than the speed of
light and in the weak field limit, one may legitimately apply
the Newtonian limit. For most galaxies these two conditions are met,
and therefore Newtonian considerations apply.

Both GR and its Newtonian limit, have
been successfully and directly tested at scales not much larger than
the
Solar System (See, e.g., Will \cite{Will93}).
 However, when one tries to apply them to galaxies or
even larger systems, the predicted behavior is usually found to be
 quite
different from what is observed.  In fact, in order to accommodate
the observations, it is customary to assume the presence of a large
amount of non-visible matter, the so-called {\it dark matter}.
 The needed amount of dark matter has to be somewhere between 90 and
  99 per cent of
the total mass of the Universe; furthermore, in order to be consistent
with the predictions derived from standard Big Bang nucleosynthesis,
it must be non-baryonic. Of course this discrepancy is commonly known
as the
{\it Dark Matter Problem}.

However, Newton's law of gravity is {\it just} a phenomenological
law that was designed by Newton to explain gravitational
 dynamics within Solar-system scales. On the other hand,
General Relativity (GR) was
developed by Einstein with the constraint in mind of recovering the
Newtonian potential in the limits of weak fields and small
(compared with the speed of light)
velocities, assuming that the phenomenological law discovered
by Newton for the Solar system could be extrapolated to
distances up to {\it infinity}. These considerations leave open the
possibility that,
at least while the {\it Dark Matter} component remains
unidentified, the possibility exists that GR,
 despite its conceptual beauty, would have to be modified
 in some way, perhaps in the same spirit as it was used to modify
Newtonian gravity for strong fields and large velocities  or,
perhaps, in other ways.

In this paper we study the problem of the rotation curves of
spiral galaxies, a case in point. When we apply the Newtonian
approximation to these
systems we find that their rotation curves should fall for large
radius as $v^2 \propto r^{-1}$, i.e., in a Keplerian fall-off.
Instead, the observed rotation velocity is
typically seen to remain constant after attaining a maximum value,
 as if
 it were to go to some asymptotic value, different for each galaxy.
This is usually explained assuming a halo of {\it Dark matter}
surrounding the visible galaxy, with the adequate shape for
accommodating the observed rotation curve.

We have posed the following (somewhat longish) question: ``{\it Is it
possible to
find a phenomenological universal Newton-like law  that can explain
the observed dynamics in spiral galaxies, without having to assume
the presence of an undetected mass component, and whose short
distance limit is compatible with the Newtonian law?}" We will offer
a positive answer to this question.

This is not the first time such an approach is taken in the
literature. Work along these lines has already been
done (for example by Kuhn \& Kruglyak \cite{KK87} and Tohline
\cite{Tohline82}),
and assumes a specific functional form for the generalized force.
This form is parametrized by some free parameters, and the
evaluation of the predicted rotation curves for some known
spiral galaxies through the corresponding numerical integrals
leads, when compared with observations, to specific values for these
parameters. This is a very interesting and direct approach, but
its success obviously depends on {\it a good choice} for the
initial of for the ``improved'' force.

We will present here a procedure that follows the inverse methodology:
we will write down an equation such that, once we know the observed
rotation
velocity of a galaxy  we readily obtain which is the force,
if any, that is able to generate that rotation curve
{\it without assuming}
the presence of any dark matter. In this way we will not have
to assume a form for the phenomenological law, we will {\it infer it}
directly from the observational data. The observed data is our
starting point, not the final result of
some ``fit". And, what could be more interesting, the method
presented here
can be, in principle, equally useful for discarding a non-Newtonian
law of
gravity as for proving its existence. Once we have the equation that
allows us to find the force from the {\it observed} velocity, we will
 apply
it to a sample of spiral galaxies and  check if there exists a common
phenomenological law that works for {\it all} the galaxies in the
sample.

We write the gravitational field as a generalization of both, the
Newtonian potential and the Newtonian force. This we do by introducing
 two functions $g(r)$ and $g_{\rm eff}(r)$ defined as:

  \begin{equation}
	\phi (r) \equiv - \frac{G_0 m_1 m_2}{r}g(r).
  \label{punctpot}
  \end{equation}
  \begin{equation}
	\vec{F}(r) \equiv - \frac{G_0 m_1 m_2}{r^2}g_{\rm eff}(r)
 \frac{\vec{r}}{r}
  \label{punctforce}
  \end{equation}
where $\phi (r)$ and $\vec{F}(r)$ are the potential and the force
experienced by two {\it point-like} particles of masses $m_1$ and $m_2$
separated by a distance $r$ and $G_0$ is Newton's constant.
The two functions $g(r)$ and $g_{\rm eff}(r)$ are related through:

  \begin{equation}
	g_{\rm eff}(r) \equiv g(r) - r g'(r),
  \label{ggef}
  \end{equation}
where the prime denotes a derivative with respect to the argument of
the function.

In order to calculate the field due to a given mass distribution
$\Omega$ described by a density function $\rho(r)$, we must
first integrate
the microscopic field over the distribution, i.e., we must perform the
integral
  \begin{equation}
	\Phi(\vec{R}) = - G_0 \int \int \int_{\Omega} d^3\vec{r} \  \
	\frac{g(|\vec{R}-\vec{r}|)}{|\vec{R}-\vec{r}|} \ \
\rho(\vec{r}),
  \label{potomeg}
  \end{equation}
which gives the potential experienced by a point mass at a
position $\vec{R}$ from the center of $\Omega$.
For a symmetric distribution (spherical or disk when considering
the disk plane) it is convenient to introduce the following
notation:
  \begin{equation}
	\Phi(R) \ \equiv \ - \ \frac{G_0 M_{\rm tot}}{R} \ \Psi(R).
  \label{potpsi}
  \end{equation}

  \begin{equation}
	V^2_{\rm rot}(R) \ \equiv \ \frac{G_0 M_{\rm tot}}{R} \
\Psi_{\rm eff}(R)
  \label{v2psi}
  \end{equation}
where $V_{\rm rot}(R)$ is the rotation velocity of a test particle
describing
circular orbits in the gravitational field generated by the
distribution $\Omega$.

It can be readily checked that the auxiliary functions
$\Psi_{\rm eff}$
and $\Psi$ satisfy the following
functional relationship:

  \begin{equation}
	\Psi_{\rm eff}(R) = \Psi(R) -R \Psi'(R).
  \label{psiphi}
  \end{equation}

As we have pointed out above, the luminous matter in many spiral
galaxies
can be well modelled by a thin disk distribution with an exponentially
decaying density (see Freeman \cite{Freeman70}).
In cylindrical coordinates this may be written as:
\begin{equation}
	\rho(\vec{r}) \equiv \rho_0 \ e^{-\alpha r} \delta(z)
\label{density}
\end{equation}
where $\alpha$ is obtained from the luminosity profile for each galaxy.

Our problem can now be paraphrased as follows:
 ``knowing the rotation velocity (i.e., $\Psi_{\rm eff}(R)$ up
to a normalization constant proportional to the mass of the galaxy),
what is the microscopic law of gravity, i.e., $g(r)$ or
$g_{\rm eff}(r)$, capable of generating that velocity field in a thin
disk galaxy?''

This problem can be solved exactly for a {\it spherical} galaxy with an
exponentially decreasing density. Here the solution can be
summarized as:
\begin{equation}
g(r)= \Psi(r) -\frac{2}{\alpha^2}\Psi''(r) +
\frac{1}{\alpha^4}\Psi^{(iv)}(r).
\label{spherical}
\end{equation}

We will consider this case in detail in a separate publication
(See Rodrigo-Blanco \cite{Rodrigo96a}).
 The line of reasoning leading to
 the proof of Eq. (\ref{spherical}) is as
follows: plug Eq.
(\ref{density}) into (\ref{potomeg}), and use the Fourier transform
of $g(r)$ together with
the addition theorem of Bessel functions to decouple
the integration variables in the integrals. Once this is done, and
after introducing the function
$\Psi(R)$, integration by parts yields Eq. (\ref{spherical}).

In the thin disk case the problem cannot be solved exactly.
For this reason
we will use an approximation that we call ``{\it Gaussian
approximation}" (we will see that, in the Newtonian case,
 this approximation is equivalent to using Gauss' law for
calculating the gravitational field and hence the name).
It should be noted that this approximation improves when one
considers
 a $g_{\rm eff}(r)$ which is an increasing function of
$r$ (which, of course, is a welcome bonus for understanding
 the rotation curves of galaxies).

In this approximation the appropiate $g_{\rm eff}(r)$ turns out to be:

	\begin{equation}
		g_{\rm eff}(x)=  \Psi_{\rm eff}(x)-\frac{1}{\alpha^2}
		\Psi_{\rm eff}''(x)
		+ \frac{2}{\alpha^2 x}\Psi_{\rm eff}'(x)
	\label{geffpsieff}
	\end{equation}
where $\Psi_{\rm eff}(x)$ has the following behavior
at the origin:

	\begin{equation}
		\Psi_{\rm eff}(0) \ = \ \Psi_{\rm eff}'(0) \ = \ 0.
	\label{geffpsieffcond}
	\end{equation}

\begin{figure*}
\epsfxsize=10cm
\epsfbox{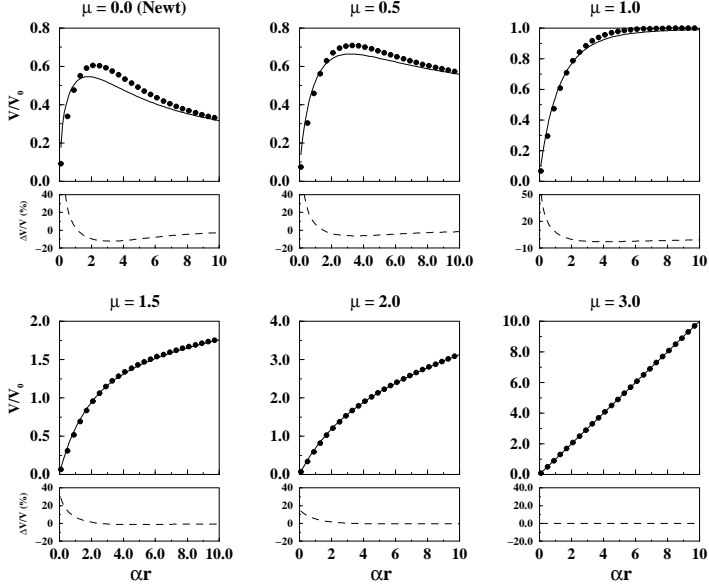}
\caption{Exact rotation curves obtained by performing a numerical
integration of the forces over a flat disk and obtaining the
rotation velocity (full points) and {\it Gaussian}
approximation (solid line) for $g_{\rm eff}(r) \propto r^{\mu}$ for
some values of $\mu$ ($\mu = 0$ is the Newtonian case).
In every case, for the sake of clarity, the
velocities are normalized by dividing by the appropriate constant:
$V_{\rm 0} \equiv \frac{G_{\rm 0}M_{\rm tot}\alpha}{(\alpha a)^{\mu}}$.
In the inset
graphs (dashed lines) we have plotted for each case the percentage of
 error made when we use the Gaussian approximation instead
of the numerical integrals, as a function of $r$.}
\label{fig:comparison}
\end{figure*}

\begin{figure*}
\epsfxsize=10cm
\epsfbox{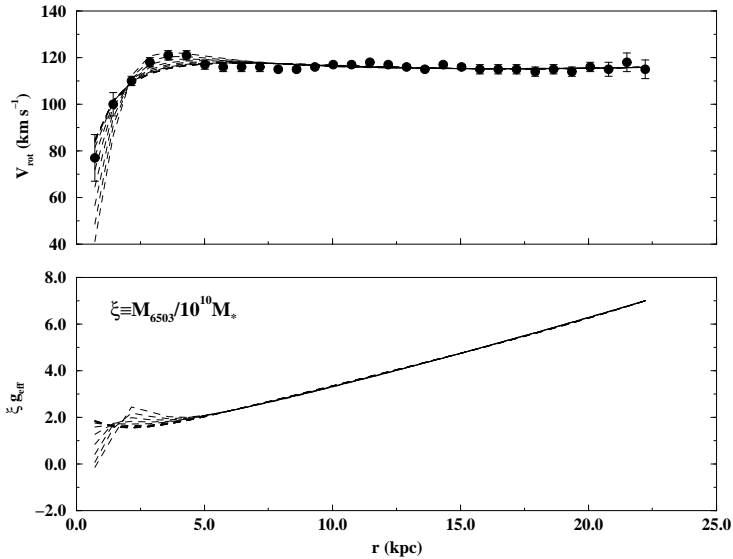}
\caption{Fits of the rotation curve of NGC 6503 using ten
different functional forms (We have used the
class of functions $v_1$, defined in Eq. (13), with a third-degree
polynomial and twelve different
 values of
$\mu$ going from $1.0$ to $2.2$ (upper graph) and the
$g_{\rm eff}(r)$
corresponding to each fit (lower graph).}
\label{fig:manyv}
\end{figure*}

The mathematical formalism applied to obtain Eqs. (\ref{geffpsieff})
and (\ref{geffpsieffcond}) is very similar to the one used for a
spherical distribution. In both cases, the use of the addition theorem
of Bessel functions leads to an infinite series of terms involving
Bessel functions of the form $J_{\rm 2k+1/2}$ and we truncate the
series keeping only the term with $k=0$.
Actually, in the presence of spherical symmetry
this is the only term that contributes to the integrals, and therefore
 the result is exact. In the thin-disk case it can be shown that
this term
dominates the integrals in the cases of interest. The mathematical
details will be given in a separate publication (See Rodrigo-Blanco
\cite{Rodrigo96b}).
 Here we give a qualitative {\it a posteriori} justification of the
goodness of the approximation. First, it can be seen that the solution
 to Eq.
(\ref{geffpsieff}), when $g_{\rm eff}(r)=1$, (Newtonian limit) is
\begin{equation}
	V^2_{\rm rot}(R) = \frac{G_0M_{\rm tot}}{R}
			[1-(1+\alpha R) e^{-\alpha R}]
		     = \frac{G_0M(R)}{R},
\label{v2g=1}
\end{equation}
where $M(R)$ is the disk mass {\it inside} a sphere\footnote{Although
 this is not the exact result, it is however what we would find if
we applied  Gauss' law as an approximation for evaluating the
gravitational field. That is why we call our
approximation {\it Gaussian}.} of radius $R$.
In order to get an idea of what happens for
 a growing $g_{\rm eff}$, let us restrict ourselves to the case when
$g_{\rm eff}(r)$
can be parametrized as a power law of the form
$g_{\rm eff}^{(\mu)}(r) \equiv (\frac{r}{a})^{\mu}$ with $\mu$
real and positive. In Fig.
(\ref{fig:comparison}) we have plotted the rotation curve obtained from
Eq. (\ref{geffpsieff}) versus the exact solution for six values of
$\mu$. It can be seen right away that,
the faster $g_{\rm eff}^{(\mu)}$ grows, the better the approximation.

Now we move on to apply our equation (Eq. (\ref{geffpsieff})) to real
galaxies. In order to do this
we have chosen a set of 9 spiral galaxies whose luminosity profiles
can be
well fitted using a thin disk model with exponential density as the
one assumed to obtain Eq. (\ref{geffpsieff}) (See refs.
Begeman \cite{BegemanTh}, Carignan \cite{Carignan85}, Persic
\& Salucci \cite{PS95}, Persic et al. \cite{PS95b} and
Mathewson et al. \cite{MFB92}). In Table
\ref{table:observed} we list the relevant observational data for these
galaxies.

The first step for applying Eq. (\ref{geffpsieff}) is to fit the
rotation velocity for each galaxy by some function, so that we
can take its derivatives. In order to do this it is important to
notice that, in the approach we are describing, there is no specific
 physical reason or prejudice for choosing one function or
another to fit the observed data: we only need that the function fits
well the observed velocities within the error bars. Apart from this
generic requirement, we
can use powers, polynomials or any other function that fits the
data so long as the corresponding velocity satisfies the following
sufficient conditions: $v(r=0)=0$, and $v(r)e^{-\alpha r} \rightarrow
0 $ as $r\rightarrow \infty$ (which are both natural conditions).
It is also important to notice that, since second derivatives of the
rotation velocity are involved in Eq. (\ref{geffpsieff}), small
differences in velocities can still lead to large differences in
$g_{\rm eff}$. This can be readily appreciated in Fig. \ref{fig:manyv},
 where we have plotted different functional fits
to the rotation velocity of NGC 6503 and the corresponding
$g_{\rm eff}$
's (up to a normalization constant).

Although all the fits are {\it statistically} acceptable, we see
qualitative differences between the different functions used to fit
and then represent the data points.
 The
differences are more significant at short distances. This range of
distances can be seen in Fig. \ref{fig:comparison} to be the one
for which our approximation is worse. Thus we will discard any values
in those regions. Actually
since these points are in any case out of the range where our method
applies, we are justified in removing them since we are mainly
interested in the
behavior of $g_{\rm eff}(r)$ for large $r$; the only restriction
is that they be compatible with $g_{\rm eff}=1$ at small distances.

\begin{figure*}
\epsfxsize=10cm
\epsfbox{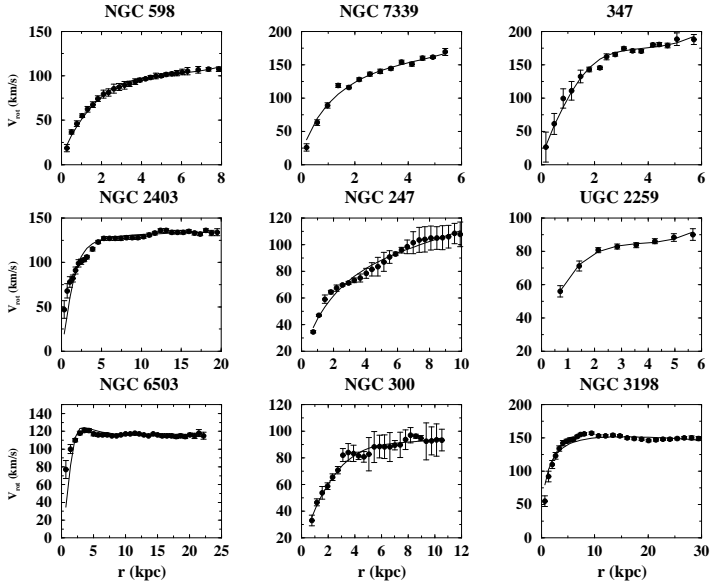}
\caption{Observed rotation curve {\it vs.} the rotation curve
{\it generated} by the $g_{\rm eff}(r)$ selected for each galaxy.}
\label{fig:fit}
\end{figure*}

\begin{table}
\caption{Relevant observed properties of selected galaxies. References
to the original data: (a) Begeman 1988;
(b) Carignan 1985; (c) Carignan et al. 1988; (d) Carignan \& Puche
1990; (e) Kent 1987; (f) Mathewson et al. 1992;
 (g) Metcalfe \& Sanks 1991; (h) Persic \& Salucci 1995, Persic et al.
 1995; (i) Puche et al. 1990}
\begin{tabular}{lllll}
\hline
Galaxy name 	& Distance 	& Scale length 	& Luminosity
	& Rot.  \\
		& (Mpc)		& (kpc$^{-1}$)	&$10^{10} L_{\sun}$
	& curve \\\hline

NGC 2403 	&3.2 $^a$	&2.1 $^a$	&0.8		$^a$
		& a  \\
NGC 3198 	&9.4 $^a$	&2.4 $^e$	&0.9		$^a$
		& a  \\
NGC 0598 	&0.9 $^g$	&1.89$^h$	&0.36		$^h$
		& h  \\
NGC 6503 	&5.9 $^a$	&1.72$^a$	&0.48 		$^a$
		& a  \\
NGC 0247 	&2.5 $^b$	&2.9 $^b$	&0.24		$^b$
		& d \\
NGC 0300 	&1.9 $^b$	&2.0 $^b$	&0.24		$^b$
		& i \\
347-g33  	&20.9$^f$	&1.46$^h$	&1.675		$^h$
		& h \\
UGC 2259 	&9.8 $^c$	&1.34$^e$	&0.1		$^c$
		& d \\
NGC 7339 	&20.6$^h$	&1.9 $^h$	&1.159		$^h$
		& h \\\hline
\end{tabular}
\label{table:observed}
\end{table}

In view of the above we use the following procedure:
({\it i}) we fit each rotation curve by a wide family of
mathematical functions, ({\it ii}) we calculate the corresponding
product
$g_{\rm eff}(r) \ M_i $ for each of those functions (where {\it i}
denotes
the particular galaxy). Thus we have a set of
$g_{\rm eff}(r)$ 's for each galaxy; furthermore, each of them can
generate
the observed rotation curve up to some multiplicative constant.
Then, ({\it iii}) we introduce all
the $g_{\rm eff}(r)$ 's in a computer program that picks up a
$g_{\rm eff}(r) \ M_i  $ for each galaxy in such a way that, once
divided
by an appropriate constant, all the $g_{\rm eff}(r)$ 's are as similar
 as
possible. For doing this, we choose a {\it standard} galaxy among
the ones in our sample (in this case we have chosen NGC 6503 because
the range of
distances for which its rotation curve is observed is the best one to
compare with the other galaxies in the sample). Then, for each galaxy,
 we fit
 the {\it mass} proportionality constant for each $g_{\rm eff}(r)$
minimizing
the $\chi^2$ of the comparison with one $g_{\rm eff}(r)$ for NGC 6503.
 Finally we
pick, for each galaxy, the $g_{\rm eff}(r)$ for which the final
$\chi^2$ is
smallest. In this way we obtain a $g_{\rm eff}(r)$ for each galaxy
that is
capable of generating its observed rotation curve within the
observational
accuracy, and we also obtain the total mass of the galaxy relative to
the mass of NGC 6503.

\begin{figure}
\epsfxsize=8.5cm
\epsfbox{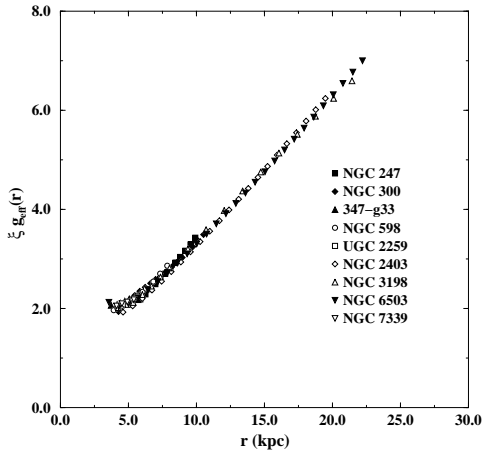}
\caption{Normalized $g_{\rm eff}(r)$  for all the galaxies in the
 sample. All of
them are multiplied by a common constant factor $\xi \equiv
M_{\rm 6503}/10^{10}M_{\sun}$.}
\label{fig:geff}
\end{figure}

Although, as mentioned before, it is irrelevant what class of
 mathematical
functions we use for the fit, it is nevertheless
interesting to mention what functions we have used here.
We have used two kinds of combinations between
powers and polynomials, labelled as $v_1$ and $v_2$, and defined by:
\begin{equation}
v_1^2(r) = \frac{r^t }{P_m(r)}
\label{v1}
\end{equation}
\begin{equation}
v_2^2(r) = r^t P_m(r)
\label{v2}
\end{equation}
where, in each case, $P_m(r)$ is a polynomial of degree $m$ in $r$ and
{\it t} is an integer greater than or equal to one.

\begin{table}
\caption{The first column indicates which kind of function $v_1$
$v_2$ (See text) was chosen to fit the observed rotation curve for
each galaxy.
The second and third columns show the parameters $\mu$ and $m$ that
better fit the data.
In the fourth column the mass corresponding to the $g_{\rm eff}(r)$
chosen for each galaxy
is given. The last
column shows the corresponding mass-to-light ratio calculated using
the observed value of
$L_{\rm B}$ this mass-to-light ratio is in units of
$M_{\sun}/L_{\rm B\sun}$ and $M_{\rm 6503}/10^{\rm 10}M_{\sun}$.}

\begin{tabular}{llllll}
\hline
 Galaxy & $\mu$ & $m$ &  & $M/M_{\rm 6503}$ & $(M/L_{\rm B})$ \\\hline
NGC 2403 & 2.1 & 2 & $V_1$ &   1.30  &    1.62\\
NGC 3198 & 1.1 & 2 & $V_1$ &   1.75  &    1.94\\
NGC 0598 & 1.2 & 3 & $V_1$ &   0.73  &    2.02\\
NGC 6503 & 2.2 & 3 & $V_1$ &   1.00  &    2.08\\
NGC 0247 & 1.1 & 2 & $V_2$ &   0.85  &    3.54\\
NGC 0300 & 1.0 & 3 & $V_1$ &   0.62  &    2.58\\
347-g33  & 1.2 & 3  & $V_1$ &  1.66  &    0.99\\
UGC 2259 & 1.0 & 2 & $V_2$ &   0.39  &    3.9\\
NGC 7339 & 1.2 & 2 & $V_2$ &   1.62  &    1.39\\\hline
\end{tabular}
\label{table:masses}
\end{table}

In Fig. \ref{fig:fit} we show the fit to the rotation curve
for each galaxy in our sample, and in Fig. \ref{fig:geff} we plot the
corresponding $g_{\rm eff}(r)$ for the galaxies multiplied by the mass
 of NGC 6503 in units of $10^{10}$ solar masses. In Table
\ref{table:masses} we list the mass of each galaxy in terms of the
mass of NGC 6503 as well as the corresponding mass to light ratio
for each galaxy. This mass-to-light ratio is in units of
$M_{\sun}/L_{\rm B\sun}$ and $M_{\rm 6503}/10^{\rm 10}M_{\sun}$.
In this table we also indicate which type of function $v_1$ or $v_2$
(See Eq. (\ref{v1}) and Eq. (\ref{v2})) was finally chosen to fit the
observed rotation
curve of each galaxy.

We end by offering some conclusions.

We have found the solution to the problem of inverting the integral
relation between an elemental (two-body) law of gravity and the
gravitational field generated by a thin disk distribution with an
exponentially decaying density. We have solved this problem in an
approximation that we have called {\it Gaussian}.
We have shown that this
approximation in general
leads to good results at large distances, although it fails at short
distances where (in any event) Many Body effects may be relevant and
overshadow the
physics of few bodies.
 This, together with the facts that rotation curves are
poorly determined in that range of distances and that the law of
gravity must be assumed to be Newton's at short scales,
allows one to discard this range in our phenomenological study.

We have selected a sample of nine galaxies such that the luminous
matter inside them can be well described by a thin
disk with exponential density. We have applied our equation to the
rotation curve of {\it each} of these galaxies and have found a law of
gravity that can generate the observed curves without the need for dark
matter (or at least, with a moderate quantity of dark matter
distributed
with the same exponential law as the luminous matter). These ``nine
laws" are statistically compatible among themselves, and point in
the direction that a single, non-Newtonian, universal (i.e. the same
for all the galaxies) law may be at
work in the realm of the galaxies.

\acknowledgements{We are grateful to Massimo Persic and Paolo Salucci
for
kindly making available some of their data prior to publication,
for their hospitality at SISSA, where a part of this work has
been done, for their helpful comments and for encouraging us
to extend the  analytical method to thin-disk distributions.
We also would like to thank Kornelis Begeman for kindly sending us a
copy
of his Ph.D. thesis.}



\newpage

\end{document}